\documentclass[review]{elsarticle}

\usepackage{lineno}

\usepackage{multirow,setspace,amssymb,amsmath,graphicx,color,rotating,subfigure}
\usepackage{lineno}
\usepackage{textcomp}
\usepackage{booktabs}    
\usepackage{CJK}
\usepackage{bm}%
\usepackage{rotating}
\usepackage{diagbox}
\usepackage{epsfig}
\usepackage{dcolumn}
\usepackage{epstopdf}
\usepackage{natbib}
\biboptions{sort&compress}
\usepackage[hidelinks]{hyperref}

\usepackage[title]{appendix}%

\usepackage[ruled,linesnumbered]{algorithm2e}

\begin{document}

\begin{frontmatter}

\author[inst1]{Su-Su Zhang}
\author[inst1]{Xiaoyan Yu}
\author[inst2,inst3]{Gui-Quan Sun\corref{ca1}}\ead{gquansun@126.com}
\author[inst1]{Chuang Liu\corref{ca1}}\ead{liuchuang@hznu.edu.cn}
\author[inst1,inst4]{Xiu-Xiu Zhan\corref{ca1}}\ead{zhanxiuxiu@hznu.edu.cn}

\title{Locating influential nodes in hypergraphs via fuzzy collective influence}

\author{\normalsize
}


\address[inst1]{Research Center for Complexity Sciences, Hangzhou Normal University, Hangzhou 311121, PR China} 
\address[inst2]{
Sino-Europe Complex Science Center, School of Mathematics, North University of China, Shanxi, Taiyuan 030051, China}
\address[inst3]{Complex Systems Research Center, Shanxi University, Shanxi, Taiyuan 030006, China}
\address[inst4]{College of Media and International Culture, Zhejiang University, Hangzhou 310058, PR China}
\cortext[ca1]{Corresponding authors.}



\begin{abstract}
Complex contagion phenomena, such as the spread of information or contagious diseases, often occur among the population due to higher-order interactions between individuals. Individuals who can be represented by nodes in a network may play different roles in the spreading process, and thus finding the most influential nodes in a network has become a crucial topic in network science for applications such as viral marketing, rumor suppression, and disease control. To solve the problem of identifying nodes that have high influence in a complex system, we propose a higher-order distance-based fuzzy centrality methods (HDF and EHDF) that are customized for a hypergraph which can characterize higher-order interactions between nodes via hyperedges. The methods we proposed assume that the influence of a node is reliant on the neighboring nodes with a certain higher-order distance. We compare the proposed methods with the baseline centrality methods to verify their effectiveness. Experimental results on six empirical hypergraphs show that the proposed methods could better identify influential nodes, especially showing plausible performance in finding the top influential nodes. Our proposed theoretical framework for identifying influential nodes could provide insights into how higher-order topological structure can be used for tasks such as vital node identification, influence maximization, and network dismantling.

\end{abstract}

\end{frontmatter}
%
%
\thispagestyle{empty}

\section{Introduction}

\makeatletter
\newcommand{\rmnum}[1]{\romannumeral #1}
\newcommand{\Rmnum}[1]{\expandafter\@slowromancap\romannumeral #1@}
\makeatother

Recently, research on measuring the influence of the nodes in a network is of theoretical and practical significance~\cite{chen2014path,bian2017identifying}, given that it has the potential to be utilized in various contexts including disease control~\cite{liu2019towards,alkhodair2020detecting}, drug targeting~\cite{harrold2013network,guo2018novel}, information dissemination~\cite{gupta2021spreading,jiang2016identifying} and network security~\cite{liu2021identifying,jiang2019identifying}. 
Classical influential node identification methods primarily concentrate on simple networks, which refer to networks composed of pairwise interactions between nodes~\cite{albert2002statistical,cimini2019statistical,boccaletti2006complex,wang2022novel}. 
However, a growing body of evidence indicates that real-world complex systems involve interactions among entities that go beyond simple pairwise relationships. A case in point is that users may form groups on social platforms to exchange emotions and ideas. Besides, multiple researchers may work together on the same project and a single drug may impact more than two proteins. The multiple interactions between entities in a complex system are usually represented by hyperedges in a hypergraph or simplicial complex~\cite{young2021hypergraph,xie2023efficient}. And in this work, we aim to characterize influential nodes on a hypergraph.

Previous researchers have conducted preliminary investigations on the identification of influential nodes in hypergraphs~\cite{li2023identifying,wang2022identifying}. Methods that consider the local or global topological structure of a hypergraph have been proposed.  For example, degree~\cite{stegehuis2021network,lu2016vital}, hyperdegree~\cite{berge1985graphs}, and hyperedge degree are methods based on the local neighbors of a node or a hyperedge. Meanwhile, vector centrality~\cite{kovalenko2022vector}, eccentricity centrality, and harmonic closeness centrality are methods based on the global structure of a hypergraph that were originally designed to evaluate the importance of a hyperedge. Recently, some work has started exploring the use of higher-order distances for the identification of important nodes, such as $s$-eccentricity centrality, $s$-harmonic closeness centrality~\cite{aksoy2020hypernetwork}, and gravity-based centrality~\cite{xie2023vital, liu2023eigenvector, tudisco2021node,zhang2022novel,curado2023novel}. While the methods mentioned above have demonstrated their efficacy in identifying essential nodes in terms of network connectivity, they have certain limitations when it comes to identifying influential nodes.

To fill this gap, we propose a higher-order distance-based fuzzy centrality to find the most influential nodes in a hypergraph. To quantify the influence of a target node, we assume that nodes that are surrounded by influential nodes are more influential. Therefore, we use a ball that is centered at the target node and with a radius determined by a higher-order distance to determine the number of surrounding nodes of the target node. Furthermore, the influence of the target nodes is computed by collecting the influence of nodes inside the ball using fuzzy sets~\cite{zadeh1997toward,parand2016combining} and Shannon entropy~\cite{shannon1948mathematical}. Experimental results demonstrate that the method we proposed can accurately identify influential nodes compared to state-of-the-art baselines.

The rest of this paper is structured as follows: in Section \textbf{2}, we provide the definition of a hypergraph and the higher-order distance on a hypergraph. In Section \textbf{3}, we provide a comprehensive explanation of the intricate steps involved in the higher-order distance-based fuzzy centrality. Additionally, we demonstrate the use of an SIR model to accurately represent the actual influence of a node. In Section \textbf{4}, we present the baselines and provide an overview of the datasets. Furthermore, the effectiveness of the proposed method is tested in Section \textbf{5}. We highlight the theoretical and practical implications to conclude the paper in Section \textbf{6} and Section \textbf{7}.


\section{Preliminary definition}\label{Definition}
\subsection{Definition of a hypergraph}
An unweighted and undirected hypergraph $H=(V, E)$ contains a node set $V = \{v_1, v_2, v_3,\cdots,v_N\}$ and a hyperedge set $E=\{e_1, e_2, e_3, \cdots, e_M\}$, where a hyperedge implies interactions between multiple nodes. 
 We construct an incidence matrix $I$ to represent the relationship between nodes and hyperedges, i.e., if node $v_i$ belongs to a hyperedge $e_j$, then $I_{ij}=1$; otherwise, it is set to 0. Mathematically speaking, it can be expressed as follows:
\begin{equation}
I_{ij}=\left\{
         \begin{array}{lr}
         1\quad \mbox{if node } v_i \mbox{ belongs to hyperedge } e_j, &  \\
         0 \quad \mbox{otherwise.}
         \end{array}
\right.
\end{equation}

\begin{figure*}[!ht]
\centering
	\includegraphics[width=\linewidth]{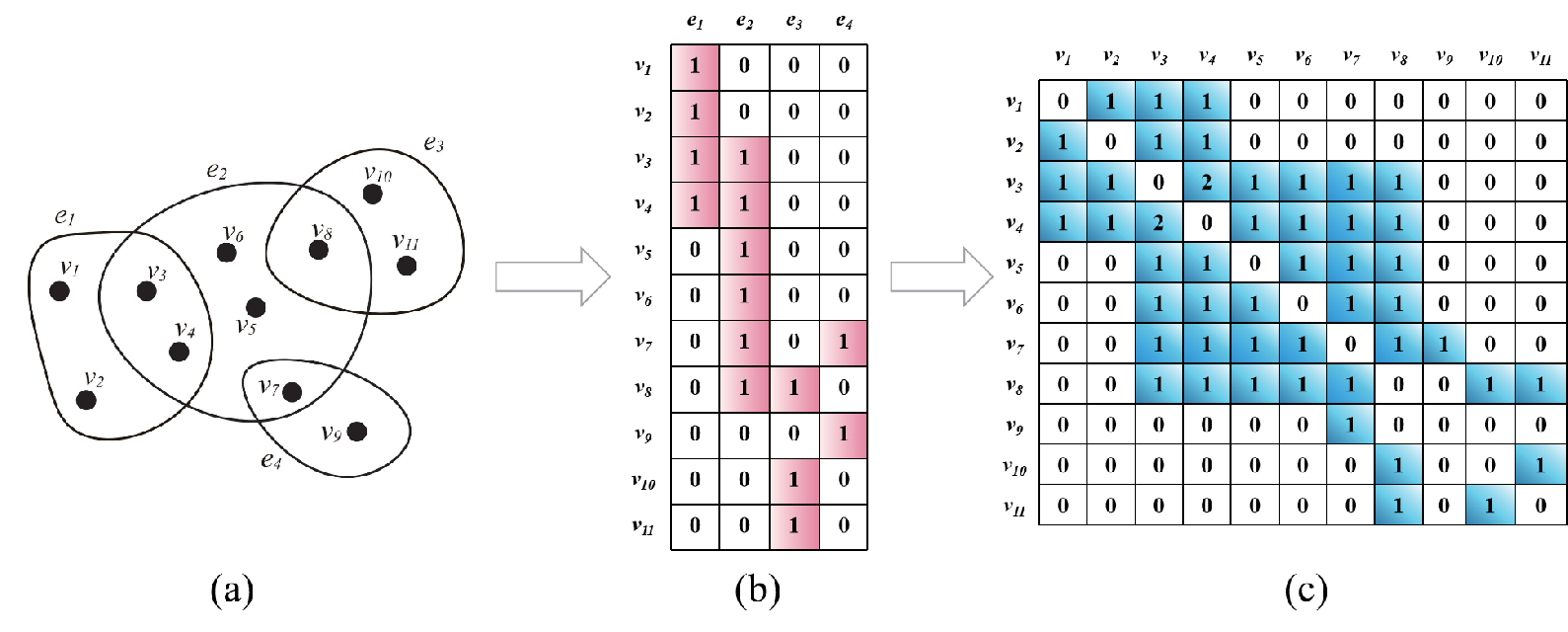}
 \caption{An illustrative example of a hypergraph: (a) a hypergraph with 11 nodes and 4 hyperedges; (b) the incidence matrix of the hypergraph given in (a); (c) the adjacency matrix of the hypergraph given in (a).}
 \label{fig: hypergraph}
\end{figure*}

Accordingly, the adjacency matrices $A$ of $H$ is obtained via the incidence matrix $I$, i.e.,
\begin{equation}
   A_{ij}=[II^T-D]_{ij},
\end{equation}
where $D$ is the diagonal matrix, $D_{ii}$ represents the number of hyperedges that node $v_i$ belongs to, and $A_{ij}$ stands for the number of hyperedges which contain both node $v_i$ and $v_j$. We show an example of a hypergraph with $11$ nodes and $4$ hyperedges in Figure~\ref{fig: hypergraph}(a), where its incidence and adjacency matrix are given in Figure~\ref{fig: hypergraph}(b) and (c).

\subsection{Definition of higher-order distance}

In a hypergraph, two hyperedges are $s$ adjacent if they share at least $s$ common nodes. An $s$-walk $w_i^l$ with a length equal to $l$ is a sequence of nodes~\cite{aksoy2020hypernetwork} which is expressed as follows:
\begin{equation}
   w_i^l = e_{i_0}, e_{i_1}, e_{i_2}, \cdots, e_{i_{l-1}}, e_{i_l};
\end{equation}
where $|e_{i_{j-1}}\cap{e_{i_{j}}}|\geq s$, $j=1, \cdots, l$, $s\geq 1$. Thus, an $s$-path between two hyperedges is an $s$-walk with nonrepeated nodes, and the $s$-distance ${d_s^e(p, q)}$ between hyperedges $e_p$ and $e_q$ is given by the length of the shortest $s$-path between them. In particular, the distance is $\infty$ if there is no $s$-path between two hyperedges.

We define the distance between two nodes on the basis of the distance between two hyperedges. Suppose that nodes $v_i$ and $v_j$ belong to hyperedges $e_p$ and $e_q$, the $s$-distance between $v_i$ and $v_j$ is denoted as $d_{s}^v(i, j)$ and can be mathematically expressed as:
\begin{equation}
d_{s}^v(i, j)=\left\{
             \begin{array}{lr}
             1\quad \quad \quad \quad \quad \quad\mbox{if } e_p = e_q&  \\
             d_{s}^e(p, q)+1 \quad  \mbox{  otherwise.}
             \end{array}
\right.
\end{equation}
\section{Method}

In this section, we will first introduce the proposed method, higher-order distance-based fuzzy centrality, for the assessment of influential nodes. The proposed method is based on the assumption that a node's influence is based on the collective influence of its higher-order neighbors, and we utilize the fuzzy sets and Shannon entropy to define the proposed centrality. Later on, we propose to use the SIR model in a hypergraph to quantify a node's real spread ability, which will be used to evaluate the effectiveness of our method.


\subsection{Higher-order distance-based fuzzy centrality}
For a central node $v_i$, we assume that the nodes close to it can influence it more. Therefore, we use fuzzy sets to define the influence of a node that is at $s$-distance $l_i^s$ on $v_i$ as

\begin{equation}
   X(l_i^s)= exp(-\frac{(l_i^s)^2}{(L^{s}_i)^2}),
\end{equation}
where $l_i^s$ represents the $s$-distance from the center node $v_i$, and $L^{s}_i$ is the radius of the $Ball(i,L^{s}_i)$ we consider and is given by 

\begin{equation}
   L^{s}_i= \lceil \frac{z_i^s}{r} \rceil,
\end{equation}
where $z_i^s$ is the maximum $s$-distance from node $v_i$ to other nodes and $r$ is a tunable parameter. By adjusting $r$, we can determine the number of nodes inside $Ball(i, L^{s}_i)$ that will influence $v_i$, i.e., with a smaller value of $r$ indicating a larger radius and thus more nodes will be incorporated into the ball. The notation $\lceil \cdot \rceil$ means we round the value up.
By using the fuzzy sets, a node that is at $s$-distance $l_i^s$ to $v_i$ has $X(l_i^s)$ influence on $v_i$, and we have $\frac{1}{e} \leq X(l_i^s) < 1$. When the value of $X(l_i^s)$ tends to $1$, it indicates that nodes located at a distance of $l_i^s$ from $v_i$ have a greater impact on its influence.

Furthermore, we assume that the number of nodes with the distance of $l_i^s$ from node $v_i$ is denoted as $n(l_i^s)$. The fuzzy number of nodes at distance $l_i^s$ to the node $v_i$ is represented as 

\begin{equation}
   f(l_i^s)= n(l_i^s)X(l_i^s)
\end{equation}

Thus, we use $F(L^{s}_i)= \sum_{l_i^s=1}^{L^{s}_i}{f(l_i^s)}$ to represent the fuzzy number of nodes in $Ball(i, L^{s}_i)$. And $p(l_i^s)$ denotes the fraction of nodes whose shortest $s$-distance from node $v_i$ is $l_i^s$, which is given by 
\begin{equation}
   p(l_i^s)= \frac{1}{e}\frac{f(l_i^s)}{F(L^{s}_i)},
\end{equation}\label{prob}
in the above equation, we use $\frac{1}{e}$ as a scaling factor to refine the probabilities to the range of $[0, \frac{1}{e}]$ in order to use Shannon entropy for node influence characterization.

We use the above fuzzy sets and probability to measure the influence of node $v_i$, the equation is given as follows:
\begin{equation}
 C_{HDF}(i)=\frac{\sum_{s=1}^{s_m}C_{HDF}^{s}(i)}{s_m},
\end{equation}
where $C_{HDF}^s(i)=\sum_{l_i^s=1}^{L^{s}_i}\frac{-p(l_i^s)ln(p(l_i^s))}{(l_i^s)^2}$ is the s-distance fuzzy centrality of node $v_i$, and $s_m (1 \leq s_m \leq s_M)$ a tunable parameter. The equation shows that we comprehensively consider the local fuzzy centrality with different $s$-distance to measure the influence of a node.

\begin{figure*}[!ht]
\centering
	\includegraphics[width=\linewidth]{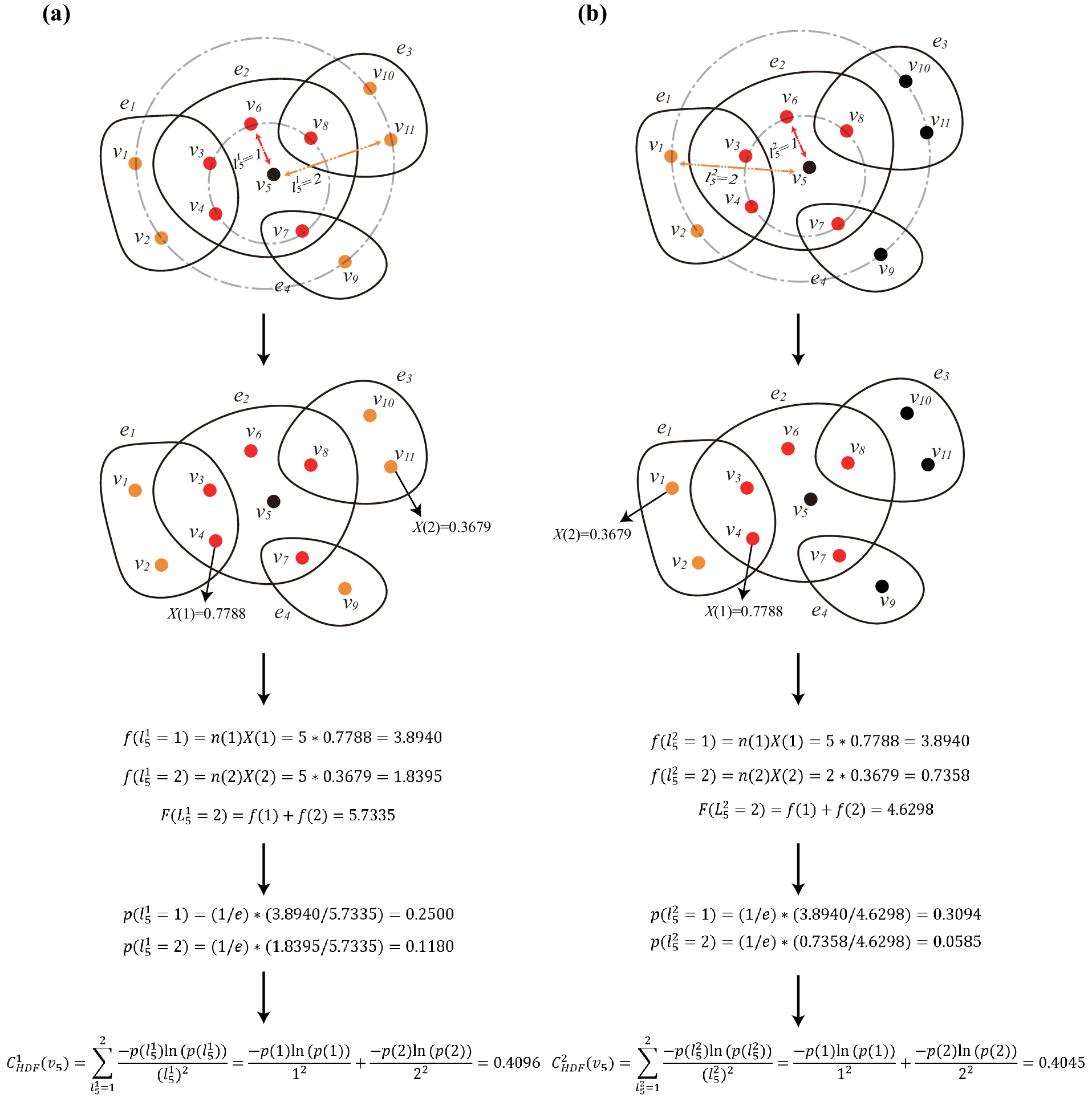}    
 \caption{A toy example of calculating higher-order distance-based fuzzy centrality.
 The target node is $v_5$ and $r=1$, $s_m=2$. The red and orange nodes represent the first and second order of neighboring nodes of  $v_5$ , respectively. We show the calculation of (a) 1-distance fuzzy centrality for node $v_5$; (b) 2-distance fuzzy centrality for node $v_5$.}
 \label{fig:HDF}
\end{figure*}

For clarity, we give a toy example of how to compute the influence of a node based on HDF, which is shown in Figure~\ref{fig:HDF}. In the figure, we set $v_5$ as the target node, and $r=1$, $s_m=2$. Figure~\ref{fig:HDF}(a) and (b) show the calculation of 1-distance and 2-distance fuzzy centrality for node $v_5$, and the values of them are $ C_{HDF}^1(v_5)=0.4096$ and $C_{HDF}^2(v_5)=0.4045$, respectively. Therefore, the final 
 HDF centrality for node $v_5$ which considers the information of different $s$-distances can be calculated as:
\begin{equation}
   C_{HDF}(v_5)=\frac{\sum_{s=1}^{2}C_{HDF}^{s}(v_5)}{2}=\frac{C^1_{HDF}(v_5)+C^2_{HDF}(v_5)}{2}=0.8141
\end{equation}



In HDF, we use the maximum s-distance from $v_i$ to other nodes, i.e., $z_i^s$, to determine the radius of the $Ball(i,L^{s}_i)$. We further incorporate all the s-distances when determining the radius, i.e., we define $L^{es}_i= \lceil \frac{\sum_{s=1}^{s_m}z_i^s}{s_m*r} \rceil$.  The other procedures are the same as HDF, and we call the new centrality method using $L^{es}_i$ as the radius as EHDF, that is, extended higher-order distance-based fuzzy centrality.


\subsection{SIR spreading model on a hypergraph}
We extend a Susceptible-Infected-Recovered (SIR) model to mimic the spreading process in a hypergraph. In the process of spreading, the nodes may be in one of the three states: Susceptible (S), Infected (I), and Recovered (R). When a node $v_i$ is infected, it will turn to the state $I$ and can randomly infect nodes that are in the same hyperedge. Every infected node has a chance of recovering to the state $R$ independently with a probability of $\mu$. Once a node is in the $R$ state, it cannot be infected by any other nodes. We note that we use $E_i$ to denote the set of hyperedges to which node $v_i$ belongs. Details of the SIR model in the hypergraph are given below.

\begin{figure*}[!ht]
\centering
	\includegraphics[width=\linewidth]{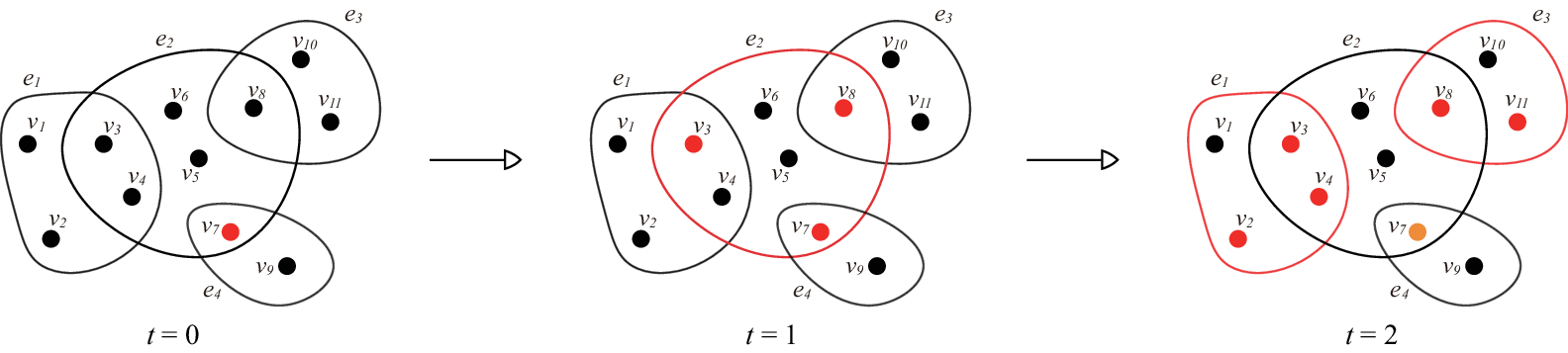}    
 \caption{An example of the SIR model on a hypergraph. The black, red, and orange nodes correspond to the S, I, and R nodes, respectively. The red hyperedge indicates that the spread is happening on it.}
 \label{fig: SIR}
\end{figure*}

\begin{itemize}

\item Initially, the seed node is labeled as the $I$ state, while the rest nodes are in the $S$ state. It is noted that no nodes are in the $R$ state at the beginning. 

\item In time step $t$, a hyperedge $e_j$ will be selected uniformly at random from $E_i$ for each infected node $v_i$. And for every $S$-state node belonging to $e_j$, it will be infected by node $v_i$ with probability $\beta$. At the same time, the nodes in the $I$ state have a probability $\mu$ to recover to the $R$ state.


\item The contagion process continues until the time step $T$.

We use the number of infected and recovered at the time step $T$ to quantify the spread capacity of the selected seed node, where $T$ is a tunable parameter. Furthermore, we set the infection probability $\beta$ slightly higher than the spreading threshold $\beta_0$ to ensure that the spreading model can spread out on a hypergraph.
\end{itemize}
We present a visual spread process of the SIR model on a hypergraph in Figure~\ref{fig: SIR}. At the initial time step $t=0$, node $v_7$ is assigned as the seed in I state.
The set of hyperedges that contains node $v_7$ is $E_7 = \{e_2,e_4\}$. In time step $t=1$, hyperedge $e_2$ is randomly selected and nodes $v_3$, $v_8$ are infected by node $v_7$, and no nodes are recovered in this step. Subsequently, infection occurs in hyperedge $e_1$ and $e_3$ at time step $t=2$, with nodes $v_2$, $v_4$ and $v_{11}$ infected and $v_7$ recovered.

\section{Baselines and Datasets}
To validate the effectiveness of the proposed methods, we introduce the state-of-the-art methods, including the family of degree centralities and centrality based on higher-order structures, as baselines. Moreover, the datasets for constructing empirical hypergraphs will be illustrated in this section as well.
\subsection{Baselines}

\textbf{Degree Centrality (DC)} measures the importance of a node by quantifying the number of  neighbors, and thus the degree of node $v_i$ is given by
\begin{equation}
   k_i=\sum_{j=1}^N{A_{ij}},
\end{equation}
where $A$ is the adjacency matrix and $N$ is the number of nodes of the hypergraph.

\textbf{Hyper Degree Centrality (HDC)} assumes that if a node belongs to more hyperedges, it has a greater chance of spreading the influence widely. The HDC of a node $v_i$ reads as
\begin{equation}
   k_i^H=\sum_{m=1}^M{I_{im}},
\end{equation}
where $I$ is the incidence matrix, and $M$ is the number of hyperedges.

\textbf{Vector Centrality (VC)} evaluates the centrality of nodes by first characterizing the centrality of the hyperedges\cite{kovalenko2022vector}. Specifically, we first project the hypergraph into a line graph\footnote{The line graph $L(H)$ is a graph of $M$ nodes. The nodes and hyperedges in $H(V, E)$ are mapped to edges and nodes in $L(H)$, respectively. That is to say, if two hyperedges $e_p$ and $e_q$ share at least one common node, there is an edge between node $v_p$ and $v_q$ in the line graph $L(H)$.}, and then calculate the eigenvector centrality values of all hyperedges. Then, the centrality value of the hyperedge will be evenly distributed to each node within the hyperedge. For a node $v_i$, suppose that the hyperedges containing $v_i$ are given by the set $E_i=\{e_{i1}, e_{i2}, \cdots, e_{iK}\}$, which will be used in the definition of the following centrality measures.
We use $c_i$ to represent the vector centrality of $v_i$, which is given by
\begin{equation}
   c_i = \sum^{K}_{k=1} c_{i k},
\end{equation}
where $c_{ik}$ is the centrality value distributed from hyperedge $e_{ik}$.

\textbf{Hyperedge Degree Centrality (HEDC):} Given the line graph of a hypergraph, we use matrix $\mathcal{A}$ to represent the adjacent relationship between two hyperedges, with $\mathcal{A}_{ij}=1$ meaning that the hyperedges $e_i$ and $e_j$ share at least one node and otherwise $\mathcal{A}_{ij}=0$~\cite{wang2010evolving,hu2021aging}. Consequently, the  degree centrality of hyperedge $e_i$ is given by 
\begin{equation}
   \mathcal{K}_i^e= \sum^{M}_{j=1} \mathcal{A}_{ij},
\end{equation}

Similarly to vector centrality, we evenly distribute the degree of the hyperedge of each hyperedge to the nodes associated with it. Given a node $v_i$ that has an associated hyperedge set $E_i=\{e_{i1}, e_{i2}, \cdots, e_{iK}\}$, the hyperedge degree centrality of $v_i$ is 

\begin{equation}
   \mathcal{K}_i = \sum_{e_{ij} \in E_i } \frac{\mathcal{K}_{ij}^e}{|e_{ij}|},
\end{equation}
where $\mathcal{K}_{ij}^e$ is the hyperedge degree centrality of $e_{ij}$.

\textbf{Eccentricity centrality (ECC)} is based on the hypothesis that highly influential hyperedges have a shorter distance from the other hyperedges.\cite{aksoy2020hypernetwork} Then, the eccentricity centrality of hyperedge $e_i$ is defined as

\begin{equation}
   {\epsilon}^e_i=\frac{1}{max_{{e_j}\in{C_1}}\{d_1^e(i,j)\}},
\end{equation}
where $C_1$ represents the $1$-connected component composed of $1$-connected hyperedges. Similarly, we distribute the centrality value of the hyperedge evenly to each node within it. Given node $v_i$ and its associated hyperedge set $E_i$, the eccentricity centrality of node $v_i$ is given by the following equation:
\begin{equation}
   {\epsilon}_i=\sum_{e_{ij} \in E_i}\frac{{\epsilon}^e_{ij}}{|e_{ij}|},
\end{equation}
where ${\epsilon}^e_{ij}$ is the eccentricity centrality of $e_{ij}$.


\textbf{Harmonic closeness centrality (HCC)} considers a hyperedge more critical if its average distance to other hyperedges is smaller\cite{aksoy2020hypernetwork}. The harmonic closeness centrality score of a hyperedge is obtained by the following equation:
\begin{equation}
   {h}^e_i=\frac{1}{M-1}\sum_{e_i,e_j\in{E},i\neq{j}}\frac{1}{d_1^e(i,j)},
\end{equation}
The harmonic closeness centrality score is evenly distributed to each node associated with it, and the harmonic closeness centrality of a node $v_i$ is defined as

\begin{equation}
   h_i=\sum_{e_{ij} \in E_i}\frac{h^e_{ij}}{|e_{ij}|},
\end{equation}
where $h^e_{ij}$ is the harmonic closeness centrality of $e_{ij}$.

\subsection{Data description}\label{Data description}
We give a detailed description of the empirical hypergraphs, which are used to evaluate the proposed method for the characterization of node influence. 
Bars-Rev and Restaurants-Rev are hypergraphs of reviews collected from \url{Yelp.com}, where hyperedges consist of users who reviewed the same bar or restaurant. Music-Rev is a hypergraph of reviews collected from Amazon. Algebra and Geometry are collected from \url{MathOverflow.net}, with users who answered the same question forming hyperedges. Email-Enron describes emails sent between employees at Enron Corporation.  The basic topological properties of these hypergraphs are presented in Table ~\ref{table: hypergraphs}, where $N$ represents the number of nodes, $M$ is the number of hyperedges, $<k>$ denotes the average degree, $<k^H>$ represents the average hyperdegree, $<k^E>$  is the average size of the hyperedges, $<l>$ and $C$ are the average shortest path length and average clustering coefficient of the corresponding simple networks, $s_M$ denotes the maximum value of the adjacency relationship in the hypergraph, and $\beta_0$ represents the infection probability of the Monte Carlo simulation we use in the experiment.
\begin{table*}[h]
    \caption{Basic topological properties of empirical hypergraphs.}
    \centering
    \resizebox{\linewidth}{!}{
    \begin{tabular}{l c c c c c c c c c } \hline\hline
     Datasets  & $N$ & $M$ & $<k>$ & $<k^H>$ & $<k^E>$ & $<l>$ & $C$ & $s_M$ & $\beta_0$ \\ \hline
      Bars-Rev & 1234  & 1194 & 174.30 & 9.62 & 9.93 & 2.1 & 0.58 & 17 & 0.016  \\ \hline
      Restaurants-Rev & 565  & 601 & 8.14 & 8.14 & 7.66 & 1.98 & 0.54 & 14 & 0.026  \\ \hline
      Music-Rev & 1106  & 694 & 167.88 & 9.49 & 15.13 & 1.99 & 0.62 & 19  & 0.012  \\ \hline
      Algebra & 423  & 1268 & 78.90 & 19.53 & 6.52 & 1.95 & 0.79 & 36 & 0.198    \\ \hline
      Geometry & 580  & 1193 & 164.79 & 21.52 & 10.47 & 1.75 &  0.82 & 63 & 0.040  \\ \hline
      Email-Enron & 143  & 1459 & 36.26 & 31.94 & 3.13 & 1.90 & 0.66 & 15 & 0.046 \\
       \hline\hline
    \end{tabular} 
    \label{table: hypergraphs}
    }
\end{table*}

\section{Experiments}\label{Experiments}
To evaluate the performance of our proposed method and the baselines, we use the Kendall correlation coefficient $\tau (\tau \in [-1,1])$ as an evaluation metric, i.e., we compute the Kendall correlation between the node rankings of a specific centrality method and the ranking of the node influence by Monte Carlo simulation by setting each node as the seed. A higher value of $\tau$ means that the centrality method can better recognize influential nodes, and vice versa. In the SIR model, we use the recovery rate as $\mu = 0.1$, and the infection probability $\beta_0$ as shown in Table~\ref{table: hypergraphs} is slightly higher than the spread threshold of each hypergraph.
 All the experiments are implemented in Python and executed independently on a server with a 2.20GHz Intel(R) Xeon(R) Silver 4114 CPU and 90GB of memory.
 
\subsection{Parametric analysis}\label{parameter}
\begin{figure*}[!ht]
\centering
	\includegraphics[width=\linewidth]{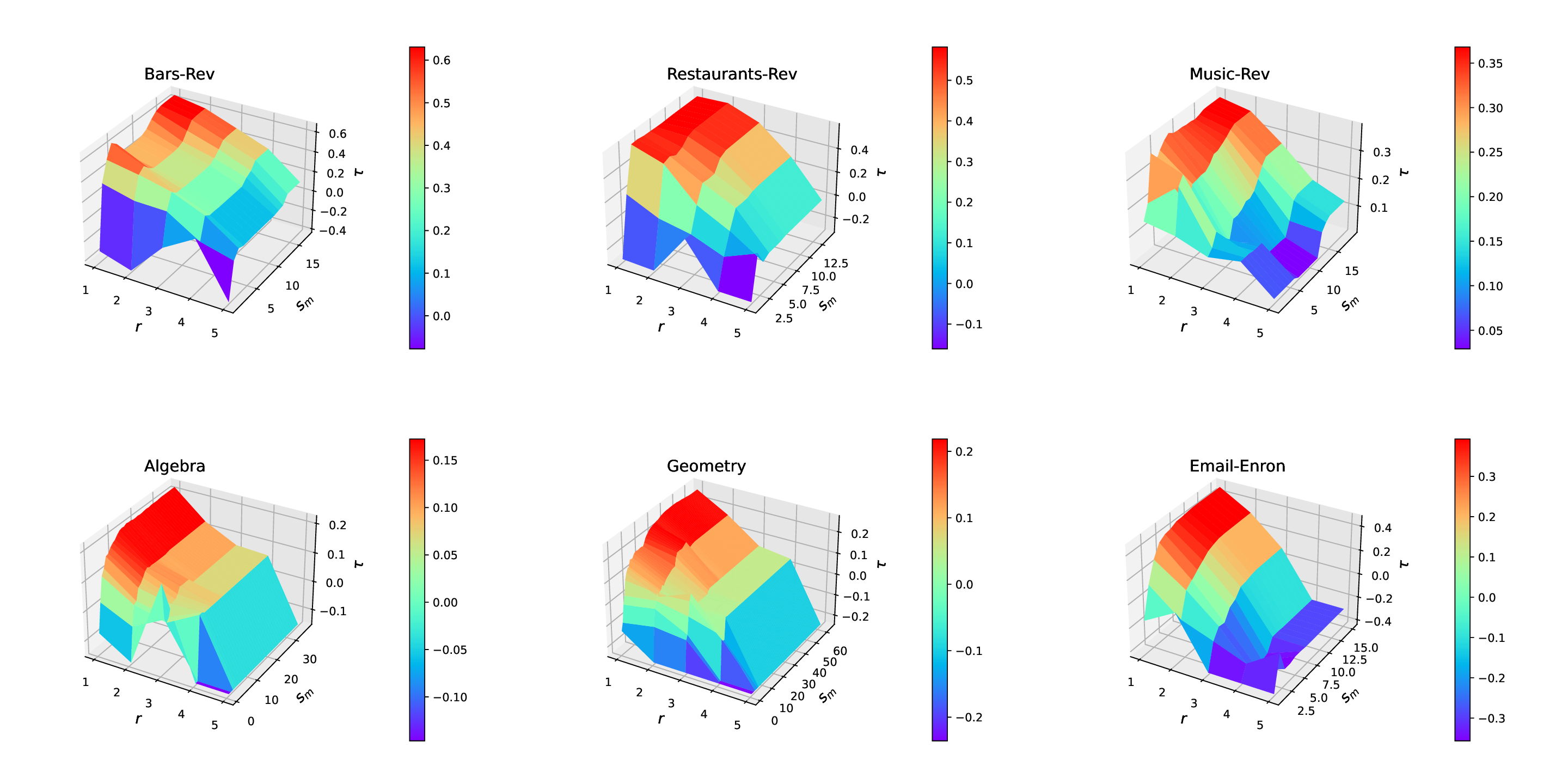}
 \caption{The change of parameters $r$ and $s_m$ on the performance of HDF in identifying influential nodes. We show the Kendall correlation coefficient for hypergraphs: Bars-Rev, Restaurants-Rev, Music-Rev, Algebra, Geometry and Email-Enron.}
 \label{fig: SS_3D}
\end{figure*}

\begin{figure*}[!ht]
\centering
	\includegraphics[width=\linewidth]{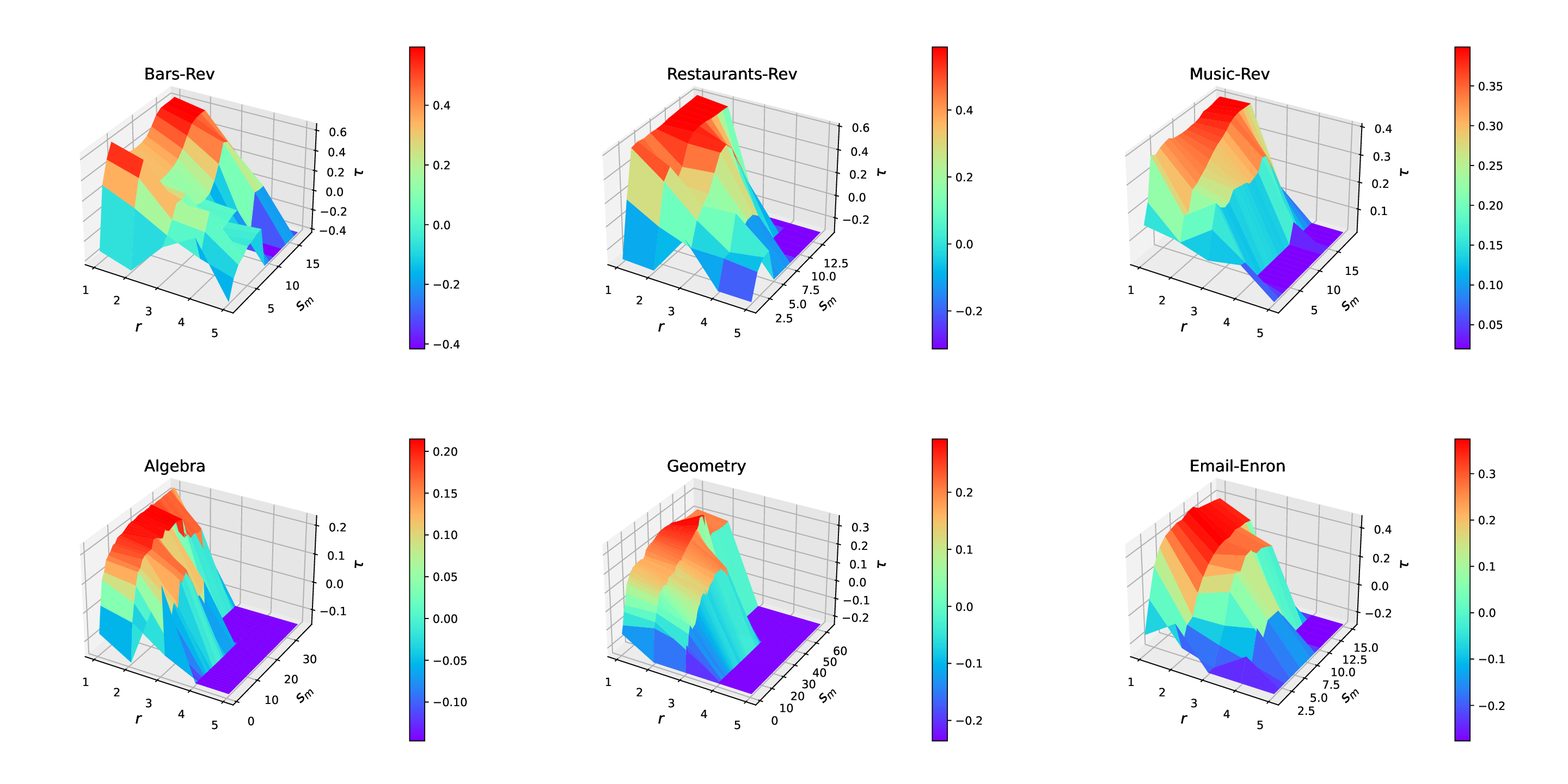}
 \caption{The change of parameters $r$ and $s_m$ on the performance of EHDF in identifying influential nodes. We show the Kendall correlation coefficient for hypergraphs: Bars-Rev, Restaurants-Rev, Music-Rev, Algebra, Geometry and Email-Enron.}
 \label{fig: DS_3D}
\end{figure*}

In HDF and EHDF, we have two parameters, namely $r$ and $s_m$, that should be adjusted to better identify the influential nodes. In Figure \ref{fig: SS_3D} and \ref{fig: DS_3D}, we display 3D heat maps to illustrate the impact of $r (r \in \{1,2,3,4,5\})$ and $s_m (1\leq s_m \leq s_M)$ on the performance of HDF and EHDF, respectively. As shown in the figure, the Kendall correlation coefficient decreases as $r$ increases, indicating that we need to consider a large radius of the ball to better identify influential nodes. In the meantime, as $s_m$ increases, the Kendall correlation coefficient initially rises and then remains roughly constant. Therefore, we choose $r=1$ (except for Geometry for EHDF, where we choose $r=2$) and $s_m=\frac{S_M}{2}$ as shown in Table~\ref{optimal_parameter} for HDF and EHDF in the following experiments.


\begin{table*}[!ht]
    \caption{The optimal parameters  $r$ and $s_m$ for the HDF and EHDF.}
    \centering
    \resizebox{\linewidth}{!}{
    \begin{tabular}{c c c c c c c c}
    \hline\hline
    \multicolumn{2}{c}{Hypergraph} & Bars-Rev & Restaurants-Rev & Music-Rev  & Algebra & Geometry &Email-Enron  \\
    \hline
    \multirow{2}*{HDF} & $s_m$ & 15 & 3 & 7  & 26 & 34 & 10\\
    \cline{2-8}
    & $r$ & 1 & 1 & 1 & 1 & 1 & 1 \\
    \hline
    \multirow{2}*{EHDF} & $s_m$ & 4 & 13 & 19  & 24 & 40 & 10\\
    \cline{2-8}
    & $r$ & 1 & 1 & 1 & 1 & 2 & 1 \\
    \hline\hline
    \end{tabular}
    \label{optimal_parameter}
    }
\end{table*}

\subsection{Performance of the proposed methods}\label{Experiments}

\begin{figure*}[!ht]
\centering
	\includegraphics[width=\linewidth]{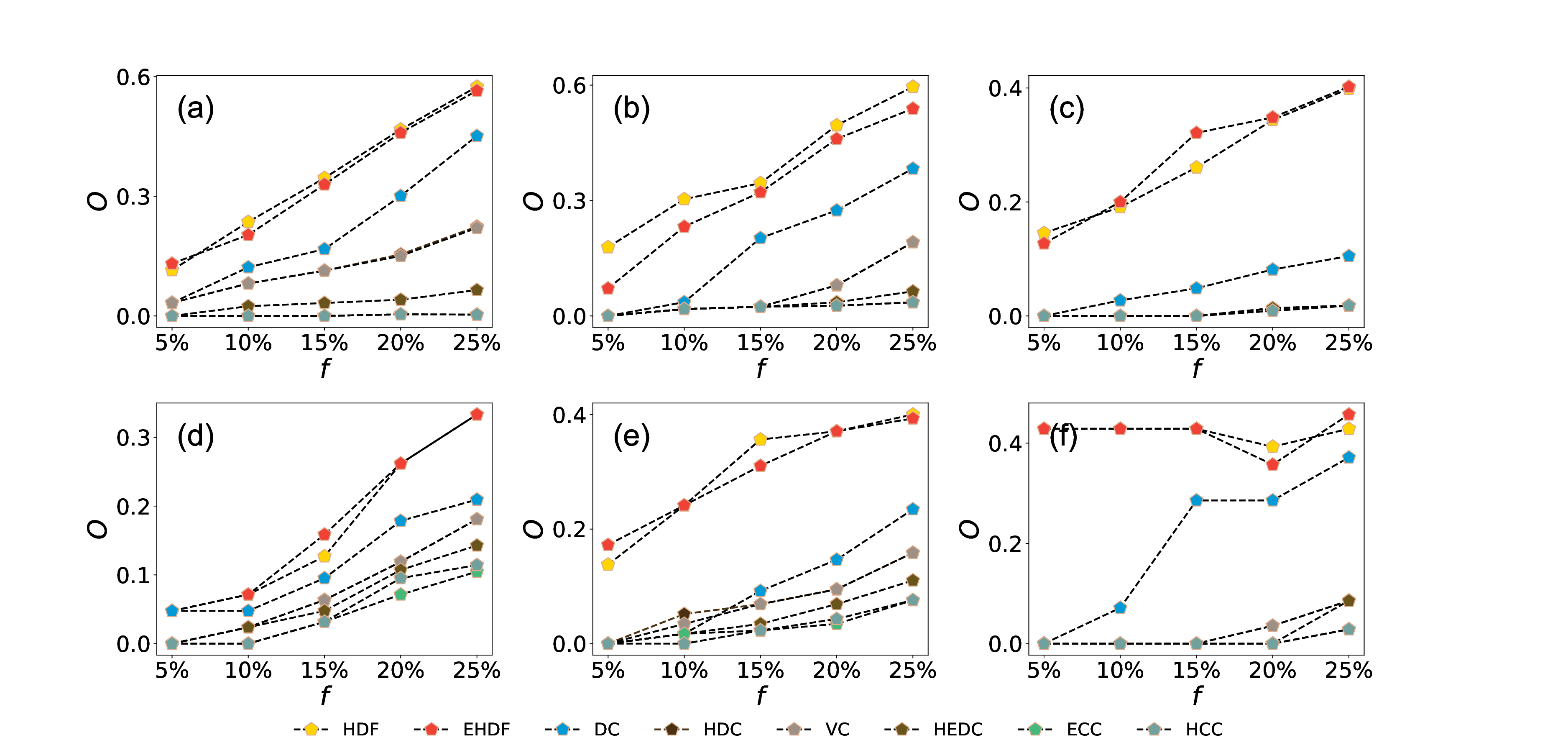}
 \caption{The overlap $O$ between the top fraction of nodes ($f$) selected by centrality methods and the top fraction of nodes ranked by Monte Carlo simulation in hypergraphs: (a) Bars-Rev; (b) Restaurants-Rev; (c) Music-Rev; (d) Algebra; (e) Geometry and (f) Email-Enron.}
 \label{fig: top_overlap}
\end{figure*}

We compare our method with the baselines to identify influential nodes, the results are given in Tables~\ref{table: Kendall_table}, \ref{table: Kendall_table_T500}  and Figure~\ref{fig: top_overlap}. In Table~\ref{table: Kendall_table} and \ref{table: Kendall_table_T500}, we show the Kendall correlation coefficient $\tau$ between the node influence ranked by a particular centrality method and Monte Carlo simulation for $T=100$ and $T=500$. The highest values of $\tau$ are highlighted in bold and the second highest values are underlined for different empirical hypergraphs.  The best performance is observed in HDF and EHDF on all hypergraphs for both $T=100$ and $T=500$, where our methods show even better performance for $T=500$. Taking Table~\ref{table: Kendall_table} as an example, EHDF outperforms in four hypergraphs, while HDF outperforms in two hypergraphs. The superiority of EHDF to HDF reveals that considering all the $s$-distances for the definition of radius can result in better identification of influential nodes.
Beyond HDF and EHDF, DC also performs well in different hypergraphs. However, the rest of the baseline methods cannot identify influential nodes, some of them even showing negative values of $\tau$. 
In reality, HDC acknowledges that nodes connected to more hyperedges are considered to have a higher degree of influence. However, the spreading model we have opted for randomly chooses a hyperedge for the contagion process. This implies that the actual influence of a node, based on its involvement in multiple hyperedges, does not necessarily hold true. Thus, the Kendall correlation coefficients for HDC and the actual spread capability of the nodes show a near-zero value in nearly all hypergraphs.
VC, HEDC, ECC, and HCC are hyperedge-based techniques that evenly distribute hyperedge centrality scores to each node within the hyperedge in order to quantify the influence of the nodes. Nevertheless, as nodes within a hyperedge may have distinct roles, the equal distribution of hyperedge centrality scores may lead to an imprecise assessment of their influence. Additionally, the baselines such as ECC and HCC only consider lower-order distance between nodes, and show even worse performance than the other baseline methods.  It should be noted that, as different values of $T$ give similar results, we thus will use $T=100$ in the following experiments.


\begin{table*}[!ht]
    \caption{The Kendall correlation coefficient $\tau$ between node influence ranked by a particular centrality method and Monte Carlo simulation. We show the results of six empirical hypergraphs: Bars-Rev, Restaurants-Rev, Music-Rev, Algebra, Geometry and Email-Enron. The best performance is shown in bold, and the second best is shown with an underline. We set $T = 100$.}
    \centering
    \resizebox{\linewidth}{!}{
    \begin{tabular}{l c c c c c c c c } \hline\hline
      Hypergraph & DC & HDC & VC & HEDC & ECC & HCC & HDF & EHDF  \\ \hline
      Bars-Rev & 0.6236 & 0.0965 & 0.0680 & -0.1783 & -0.5725 & -0.5490  & \textbf{0.6573} & \underline{0.6419} \\ \hline
      Restaurants-Rev & 0.5047 & 0.0469 & 0.0500 & -0.0648 & -0.4070 & -0.3878  & \underline{0.5851} & \textbf{0.5963} \\ \hline
      Music-Rev & 0.0836  & -0.2494 & -0.2466 & -0.2726 & -0.4711 & -0.4337  & \underline{0.3862} & \textbf{0.3974} \\ \hline
      Algebra & \underline{0.2525}  & 0.0362 & 0.03615 & 0.2115 & -0.1671 & -0.1276  & 0.2475 & \textbf{0.2556} \\ \hline
      Geometry & \underline{0.2613}  & -0.0495 & -0.0504 & -0.1644 & -0.3056 & -0.2998  & 0.2574 & \textbf{0.3368} \\ \hline
      Email-Enron & 0.4376  & -0.0920 & -0.0937 & -0.0580 & -0.1372 & -0.1110  & \textbf{0.4673} & \underline{0.4669} \\ 
       \hline\hline
    \end{tabular} 
    \label{table: Kendall_table}
    }
\end{table*}

\begin{table*}[!ht]
    \caption{The Kendall correlation coefficient $\tau$ between node influence ranked by a particular centrality method and Monte Carlo simulation. We show the results of six empirical hypergraphs: Bars-Rev, Restaurants-Rev, Music-Rev, Algebra, Geometry and Email-Enron. The best performance is shown in bold, and the second best is shown with an underline. We set $T=500$.}
    \centering
    \resizebox{\linewidth}{!}{
    \begin{tabular}{l c c c c c c c c } \hline\hline
      Hypergraph & DC & HDC & VC & HEDC & ECC & HCC & HDF & EHDF  \\ \hline
      Bars-Rev & 0.6266 & 0.0915 & 0.0624 & -0.1834 & -0.5825 & -0.5587 & \textbf{0.6623} & \underline{0.6541} \\ \hline
      Restaurants-Rev & 0.5077 & 0.0356 & 0.038 & -0.0775 & -0.4275 & -0.4057 & \underline{0.6125} & \textbf{0.6255} \\ \hline
      Music-Rev & 0.0790 & -0.2787 & -0.2756 & -0.3098 & -0.5095 & -0.472 & \underline{0.4076} & \textbf{0.4137} \\ \hline
      Algebra & 0.2361 & -0.0163 & -0.0164 & -0.027 & -0.2395 & -0.1975 & \textbf{0.2550} & \underline{0.2612} \\ \hline
      Geometry & \underline{0.2657} & -0.0563 & -0.0575 & -0.1781 & -0.3231 & -0.3164 & 0.2626 & \textbf{0.3344} \\ \hline
      Email-Enron & \underline{0.4727} & -0.0910 & -0.0923 & -0.0621 & -0.1409 & -0.1136 & \textbf{0.4809}& 0.4644 \\ 
       \hline\hline
    \end{tabular} 
    \label{table: Kendall_table_T500}
    }
\end{table*}

Table~\ref{table: Kendall_table} and \ref{table: Kendall_table_T500} show the performance of the proposed methods in ranking nodes globally. However, the top-ranked nodes are more important in reality. Therefore, we evaluate the centrality methods in identifying the top-ranked influential nodes in Figure~\ref{fig: top_overlap}, where the x-axis shows the top $f$ fraction of nodes ($f \in \{5\%, 10\%,15\%, 20\%, 25\%\}$) and the y-axis shows the overlap between the $f$ fraction of the node sequence ranked by a specific centrality method and Monte Carlo simulation. The overlap $O_{ij}$ between two sets $B_i$ and $B_j$ which have the same number of elements is defined as $O_{ij}=\frac{|B_i\cap B_j|}{|B_j|}$. 
The figures show that the top-ranked node sequences of EHDF and HDF are more overlapped with the SIR model-ranked node sequence across different hypergraphs and different fractions $f$ of nodes, further implying that our method could better identify the influential top nodes than the baselines.

\begin{figure}[!ht]
    \centering
    \includegraphics[width=\linewidth]{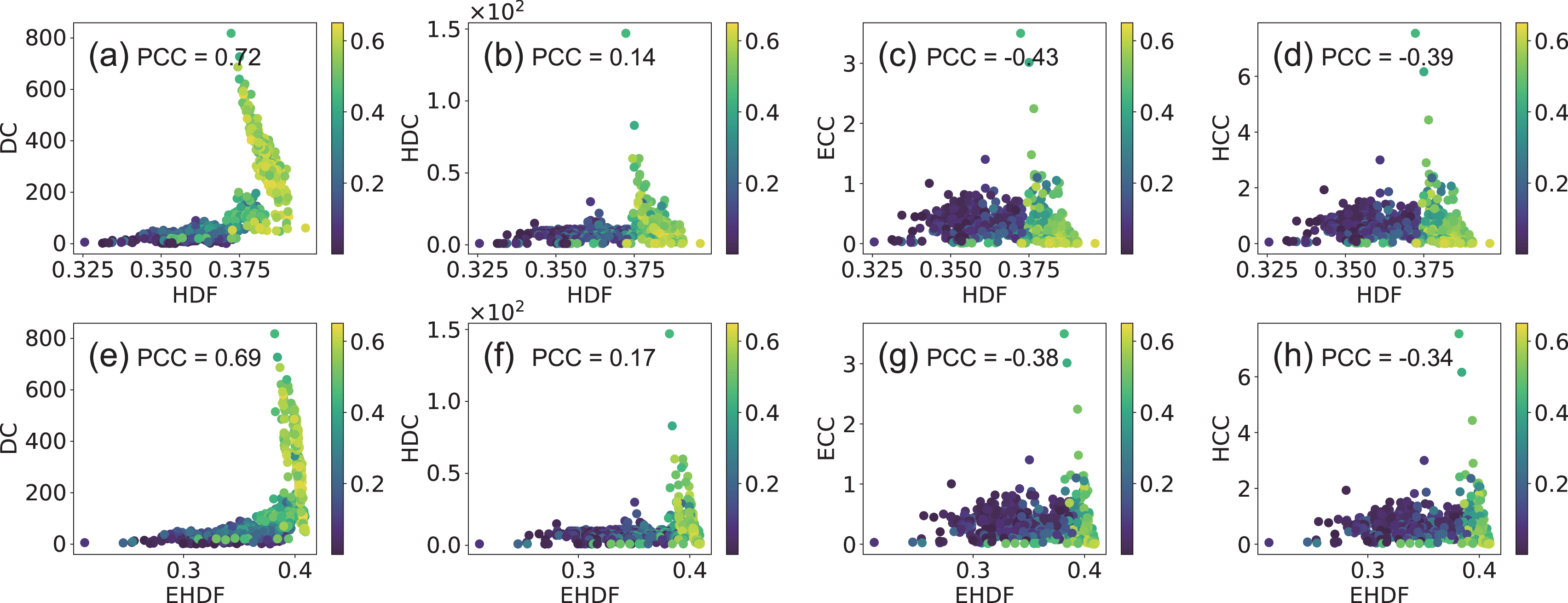}
    \caption{The correlation between the proposed methods (HDF and EDHF) and four baseline methods (DC, HDC, ECC, and HCC) in Bars-Rev. Each point in the figures shows a node in the hypergraph and the color of the points describes the spreading ability of the nodes. Figures (a-d) illustrate the correlation between HDF and four baselines. Figures (e-h) illustrate the correlation between EHDF and four baselines.
    }\label{fig:Bars-Rev_correlation}
\end{figure}

\begin{figure}[!ht]
    \centering
    \includegraphics[width=\linewidth]{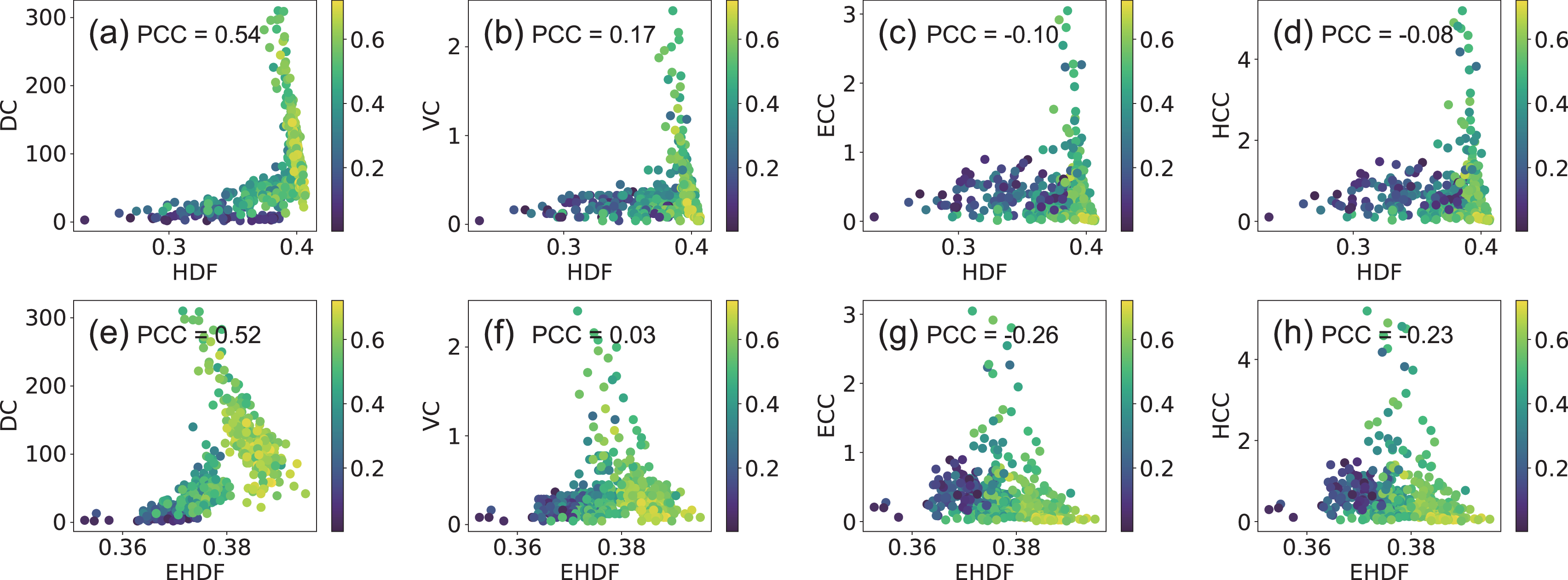}
    \caption{
    The correlation between the proposed methods (HDF and EDHF) and four baseline methods (DC, VC, ECC, and HCC) in Restaurants-Rev. Each point in the figures shows a node in the hypergraph and the color of the points describes the spreading ability of the nodes. Figures (a-d) illustrate the correlation between HDF and four baselines. Figures (e-h) illustrate the correlation between EHDF and four baselines.
    }\label{fig:Restaurants-Rev_correlation}
\end{figure}

\begin{figure}[!ht]
    \centering
    \includegraphics[width=\linewidth]{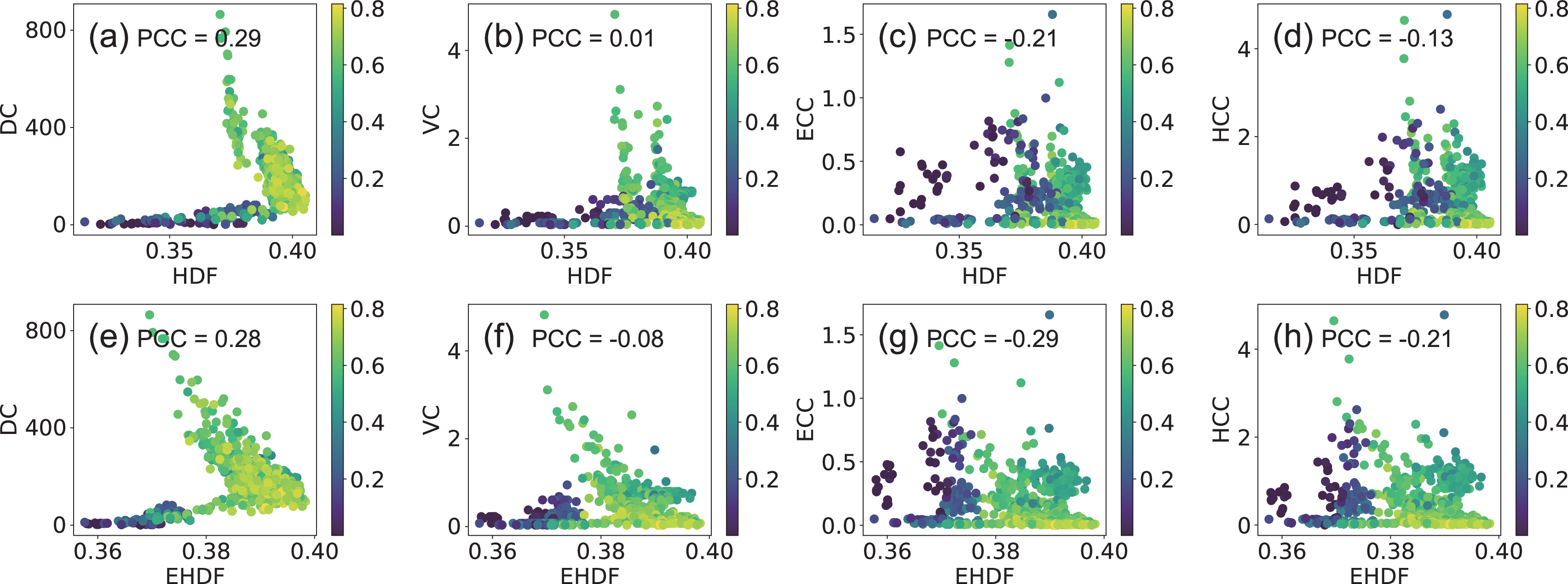}
    \caption{The correlation between the proposed methods (HDF and EDHF) and four baseline methods (DC, HDC, ECC, and HCC) in Music-Rev. Each point in the figures shows a node in the hypergraph and the color of the points describes the spreading ability of the nodes. Figures (a-d) illustrate the correlation between HDF and four baselines. Figures (e-h) illustrate the correlation between EHDF and four baselines.
    }\label{fig:Music-Rev_correlation}
\end{figure}

\begin{figure}[!ht]
    \centering
    \includegraphics[width=\linewidth]{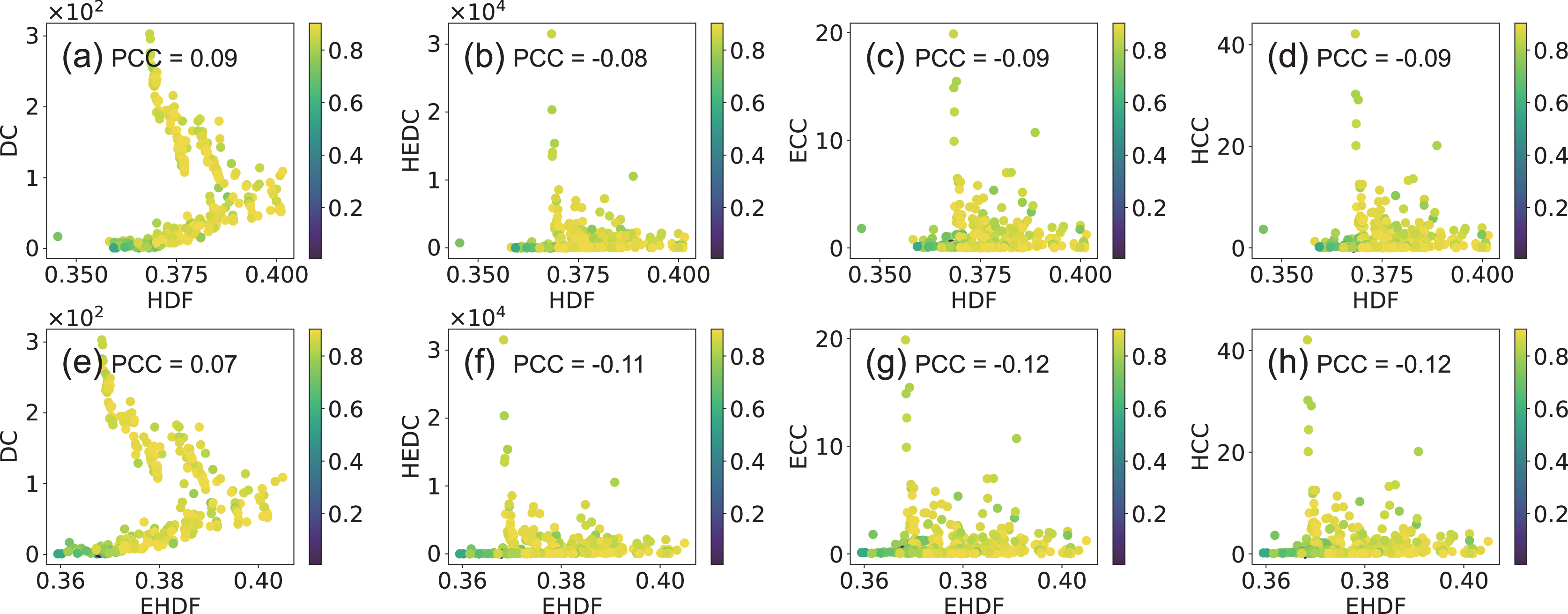}
    \caption{The correlation between the proposed methods (HDF and EDHF) and four baseline methods  (DC, HEDC, ECC, and HCC) in Algebra. Each point in the figures shows a node in the hypergraph and the color of the points describes the spreading ability of the nodes. Figures (a-d) illustrate the correlation between HDF and four baselines. Figures (e-h) illustrate the correlation between EHDF and four baselines.
    }\label{fig:Algebra_correlation}
\end{figure}

\begin{figure}[!ht]
    \centering
    \includegraphics[width=\linewidth]{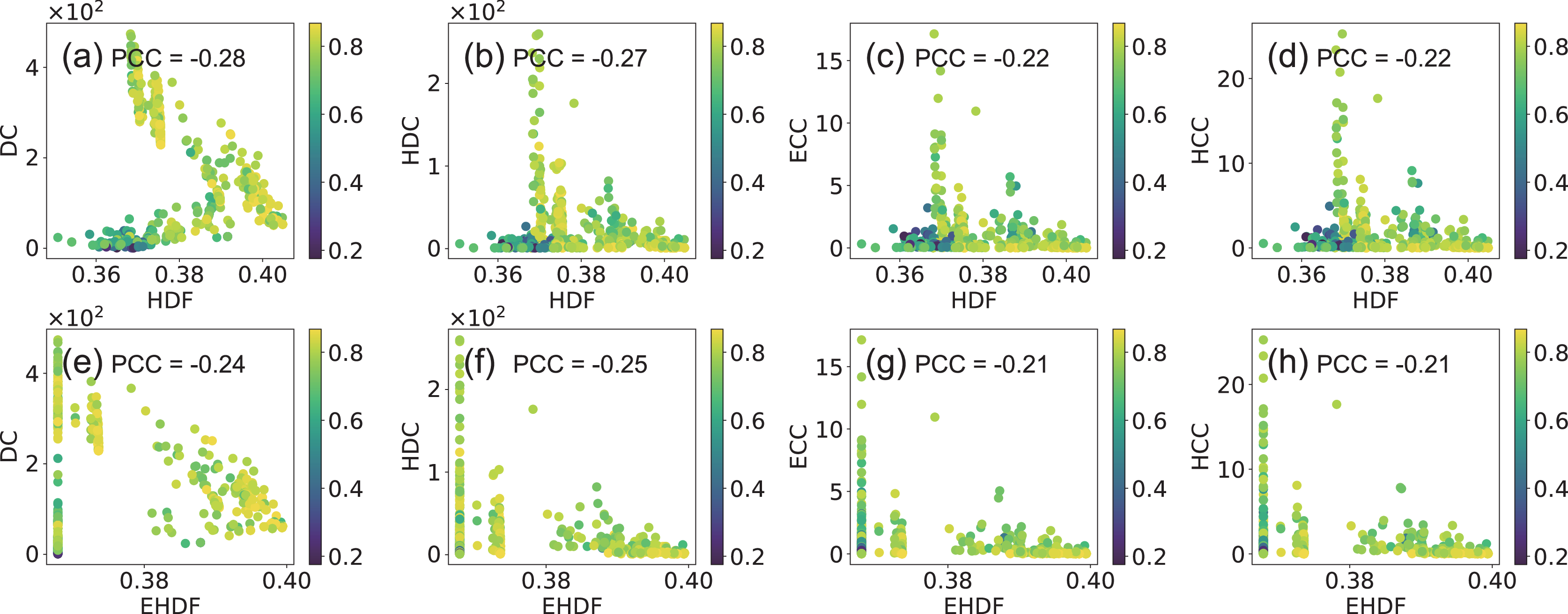}
    \caption{The correlation between the proposed methods (HDF and EDHF) and four baseline methods (DC, HDC, ECC, and HCC) in Geometry. Each point in the figures shows a node in the hypergraph and the color of the points describes the spreading ability of the nodes. Figures (a-d) illustrate the correlation between HDF and four baselines. Figures (e-h) illustrate the correlation between EHDF and four baselines.
    }\label{fig:Geometry_correlation}
\end{figure}

\begin{figure}[!ht]
    \centering
    \includegraphics[width=\linewidth]{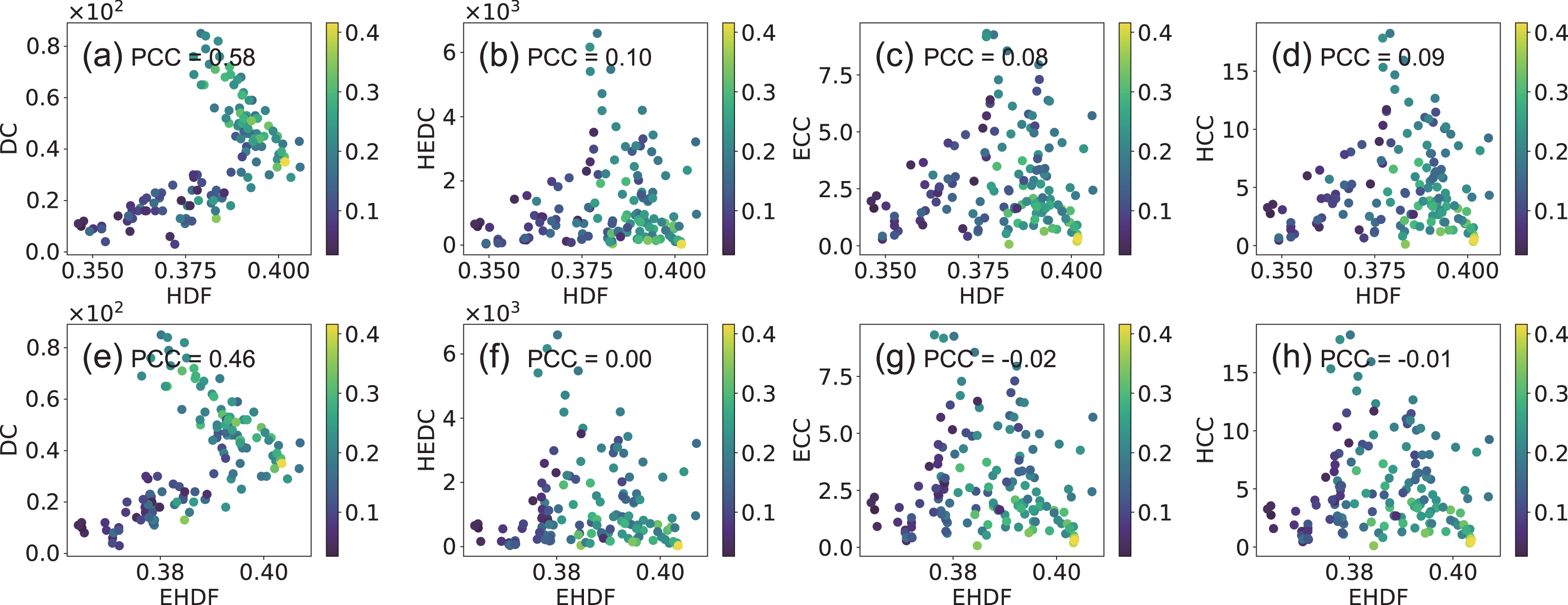}
    \caption{The correlation between the proposed methods (HDF and EDHF) and four baseline methods (DC, HeDC, ECC, and HCC) in Email-Enron. Each point in the figures shows a node in the hypergraph and the color of the points describes the spreading ability of the nodes. Figures (a-d) illustrate the correlation between HDF and four baselines. Figures (e-h) illustrate the correlation between EHDF and four baselines.
    }\label{fig:email-Enron-full_graph_correlation}
\end{figure}

To gain a deeper understanding of the reasons behind the superior performance of the proposed methods compared to the baselines, we perform a more detailed analysis of the relationship between the centrality methods, and the results are given in Figures~\ref{fig:Bars-Rev_correlation} to \ref{fig:email-Enron-full_graph_correlation}. For each hypergraph, we select the two best and two worst performing methods evaluated by the Kendall correlation coefficient given in Tables~\ref{table: Kendall_table} and \ref{table: Kendall_table_T500}, and show the correlation of each of these methods with HDF (Figures (a-d)) or EHDF (Figures (e-h)). Taking Bars-Rev as an example, DC and HDC perform the best apart from the methods we proposed, and ECC and HCC show the worst performance. In each of the figures, a point represents a node in the hypergraph, where the color of the points indicates the spreading capacity of the nodes (determined by Monte Carlo simulation), with the color transitioning from blue to yellow to indicate nodes with low to high influence. According to Figures~\ref{fig:Bars-Rev_correlation} to \ref{fig:email-Enron-full_graph_correlation}, we observe a consistent pattern in all hypergraphs, i.e., the actual spreading ability of the nodes aligns closely with their centrality scores calculated using HDF (or EHDF). Specifically, blue nodes are predominantly located on the left side (corresponding to low HDF or EHDF values), while yellow nodes are predominantly located on the right side (corresponding to high HDF or EHDF values). However, the centrality values of the nodes in the baselines are not consistent with the real spread capacity of the nodes, which is in line with their bad performance illustrated in Tables~\ref{table: Kendall_table}, \ref{table: Kendall_table_T500} and Figure~\ref{fig: top_overlap}. For example, nodes with degrees $50$ to $800$ share close values of spread ability in Bars-Rev. We further give the Pearson correlation coefficient (PCC) between the node sequence ranked HDF (EHDF) and the selected baselines, as shown in Figures~\ref{fig:Bars-Rev_correlation} to \ref{fig:email-Enron-full_graph_correlation}. We observe that the proposed methods show high PCC values with the two best baselines and low PCC values with the worst ones.

\begin{figure}[!ht]
    \centering
    \includegraphics[width=\linewidth]{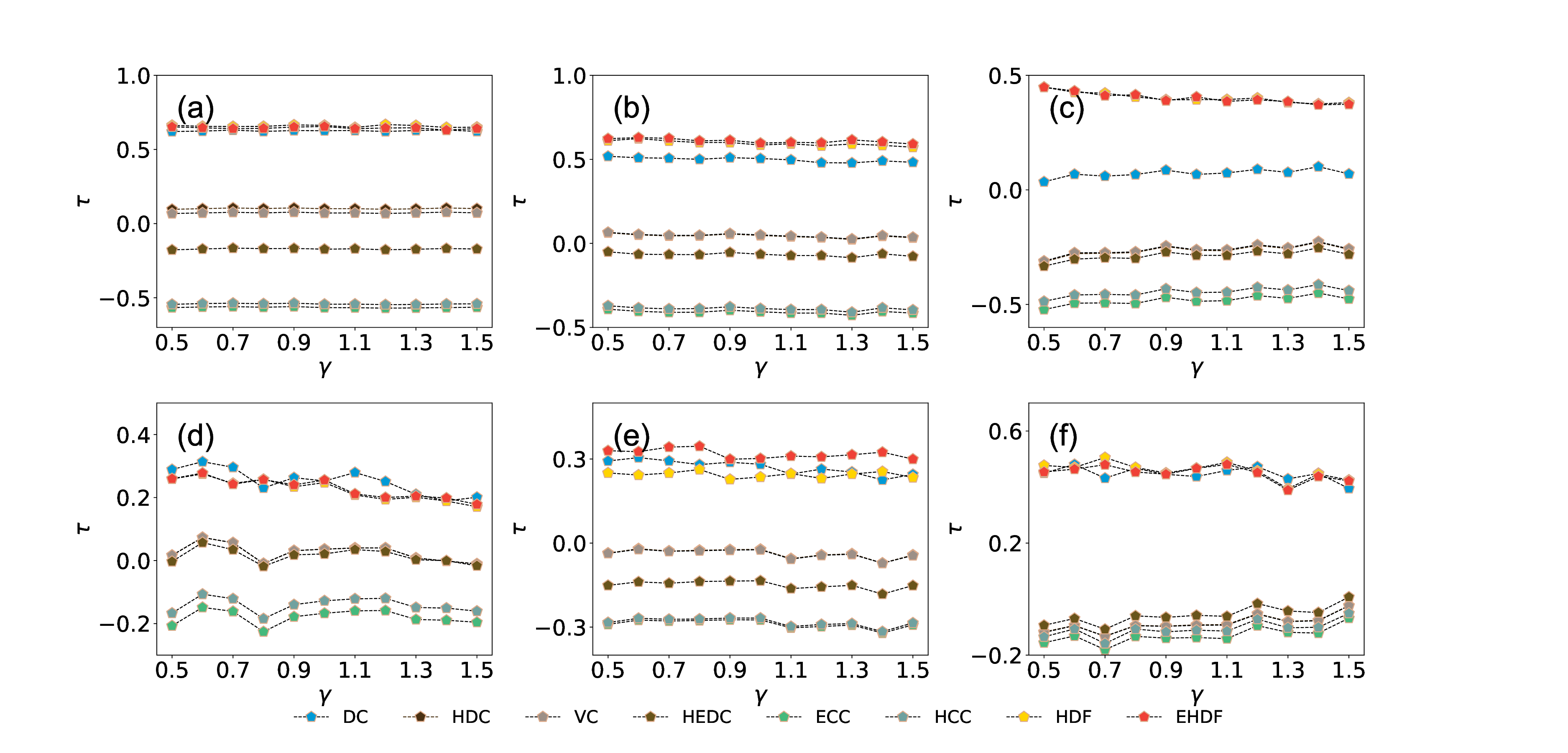}
    \caption{Kendall correlation coefficient $\tau$ between centrality scores obtained from various methods and the spread ability of nodes at various parameter $\gamma$ in the six empirical hypergraphs:
     Bars-Rev, Restaurants-Rev, Music-Rev, Algebra, Geometry and Email-Enron.
    }\label{fig:sensitivity}
\end{figure}

To further evaluate the resilience of our approach, we conduct tests by modifying the infection probability $\beta=\gamma \beta_0$. The outcomes are presented in Figure \ref{fig:sensitivity}. 
We calculate the Kendall correlation coefficient $\tau$ to examine the relationship between centrality scores obtained from various methods and the spread ability of nodes. The parameter $\gamma$ ranges from 0.5 to 1.5, with an interval of 0.1.
As the value of $\gamma$ changes, HDF and EHDF consistently demonstrate stable and excellent performance. Similarly, the other baselines also maintain stable performance as $\gamma$ varies, but show worse performance than our proposed methods. In particular, DC outperforms other baseline algorithms, especially in the Algebra network, where DC exhibits almost the best performance across different values of $\gamma$, yet HDF and EHDF remain highly competitive as well.

\section{Discussion}
\subsection{Findings}
In this study, we observe that the proposed approach, namely higher-order distance-based fuzzy centrality (HDF), shows outstanding effectiveness in the identification of influential nodes. The baselines are unable to accurately rank the influence of nodes or identify the most influential nodes due to their limitations, such as only considering local or low-order information of a hypergraph. In contrast, the experimental results suggest that HDF and EHDF, which incorporate higher-order information, could identify influential nodes with high accuracy.

\subsection{Theoretical contribution}
This study sheds light on the problem of locating influential nodes on hypergraphs and addresses the inadequacies in previous research regarding the utilization of higher-order information. In the design of
higher-order distance-based fuzzy centrality, the influence of a target node is dependent on the influence of neighboring nodes at different higher-order distances, and the influence is collected by a nonlinear function that incorporates fuzzy sets and Shannon entropy. The effectiveness of our approach is dependent on the utilization of an adjustable radius for the ball in order to ascertain the neighboring nodes of each target node, as well as the non-linear function. This offers a theoretical support for identifying influential nodes in higher-order networks.

\subsection{Practical significance}
The investigation of identifying high-influence nodes on hypergraphs is of great practical importance, as many real-world systems involve complex interactions among entities, represented by hyperedges. This research expands the potential applications of high-influence node identification to various domains, including viral marketing, epidemic prevention, and social opinion management.

\section{Conclusion}

The spreading dynamics on a hypergraph is usually not only through pairwise interactions, but also through higher-order interactions involving multiple nodes. In this work, we tackle the problem of identifying influential nodes in a hypergraph that could characterize higher-order interactions between nodes. we start with the definition of higher-order distance and describe the SIR spreading model, which aims to model the real influence of a node in a hypergraph. Based on the fuzzy collective influence that collects the influence of nodes inside the ball using fuzzy sets and Shannon entropy to quantify the influence of the target node, we propose a higher-order distance-based fuzzy centrality (i.e., HDF and EHDF) to solve the problem. To validate the effectiveness of our methods, we perform experiments on six empirical hypergraphs from various domains. The results show that our methods are superior to state-of-the-art benchmarks in terms of ranking influential nodes, especially in identifying the top influential nodes.

Despite the effectiveness of HDF and EHDF in identifying influential nodes, the proposed methods utilize a high-order distance between nodes, which is with high computational complexity, especially for large-scale networks. Therefore, one possible direction for future work could focus on proposing approaches to efficiently calculate the higher-order distance. Besides, the theoretical framework we have proposed for the identification of influential nodes has the potential to be applied to other types of higher-order networks, including simplicial complexes~\cite{raj2023some}, temporal hypergraphs~\cite{cencetti2021temporal}, and multilayer hypergraphs~\cite{zhen2023community}.

\section{CRediT authorship contribution statement}
\textbf{Su-Su Zhang:} Conceptualization, Methodology, Formal Analysis, Investigation, Original Draft Preparation, Review \&editing. \textbf{Xiaoyan Yu:} Investigation, Original Draft Preparation, Data Curation, Visualization. \textbf{Chuang Liu:} Conceptualization, Methodology, Supervision, Review \& editing. \textbf{Xiu-Xiu Zhan:} Conceptualization, Methodology, Validation, Investigation, Supervision, Review \& editing.

\section{Data and code availability}
Data will be available on request.

\section{Acknowledgments}
This work was supported by the Natural Science Foundation of Zhejiang Province (Grant No. LQ22F030008), the Natural Science Foundation of China (Grant No. 61873080) and the Scientific Research Foundation for Scholars of HZNU (2021QDL030).

\bibliography{Tdis}

\end{document}